\documentclass[aip,pop,floats,reprint,superscriptaddress]{revtex4-1}
\usepackage{amssymb}
\usepackage{amsmath}
\usepackage{graphicx}
\usepackage{epstopdf}
\allowdisplaybreaks
\begin{document}
 \title{Witness emittance growth caused by driver density fluctuations in plasma wakefield accelerators}
 \author{V.A.Minakov}
 \affiliation{Budker Institute of Nuclear Physics SB RAS, 630090, Novosibirsk, Russia}
 \affiliation{Novosibirsk State University, 630090, Novosibirsk, Russia}
 \author{M.Tacu}
 \affiliation{ \'Ecole Normale Sup\'erieure Paris-Saclay, 94230, Cachan, France}
 \author{A.P.Sosedkin}
 \author{K.V.Lotov}
 \affiliation{Budker Institute of Nuclear Physics SB RAS, 630090, Novosibirsk, Russia}
 \affiliation{Novosibirsk State University, 630090, Novosibirsk, Russia}
 \date{\today}
 \begin{abstract}
We discovered a novel effect that can cause witness emittance growth in plasma wakefield accelerators. The effect appears in linear or moderately nonlinear plasma waves. The witness experiences a time-varying focusing force and loses quality during the time required for the drive beam to reach transverse equilibrium with the plasma wave. The higher the witness charge, the lower the emittance growth rate because of additional focusing of the witness by its own wakefield. However, the witness head always degrades, and the boundary between degraded and intact parts gradually propagates backward along the witness bunch.
 \end{abstract}
 \maketitle

\section{Introduction}

Plasmas offer new opportunities in the area of novel acceleration techniques, which are associated with high accelerating gradients possible in the plasmas.\cite{NatPhot7-775,RAST9-19,RAST9-63,RAST9-85,RAST9-209} The goal is not only to reduce the acceleration distance, but also to preserve the beam quality. In particular, the normalized emittance of the accelerated bunch (witness) must be conserved or grow insignificantly in the plasma. Several sources of emittance growth have already been identified: motion of plasma ions,\cite{PRL95-195002} multiple Coulomb scattering,\cite{PAC07-3097,PoP22-083101} transitions between plasma sections and conventional focusing elements,\cite{JAP112-044902,PRST-AB15-111303,PRST-AB16-011302,PRL116-124801} and misalignment of multiple plasma sections.\cite{PRST-AB3-071301,PRST-AB3-101301} Solutions to the discovered problems gradually emerge.

We have found one more effect that can cause emittance growth. It is related to non-stationarity of the drive beam. The driver needs some time to approach a transverse equilibrium with the plasma wave.\cite{PoP24-023119} The driver shape changes during equilibration, causing temporal fluctuations of the focusing force in the downstream wake, which in turn heat the witness.

\begin{figure}[b]\centering
 \includegraphics[width=0.95\columnwidth]{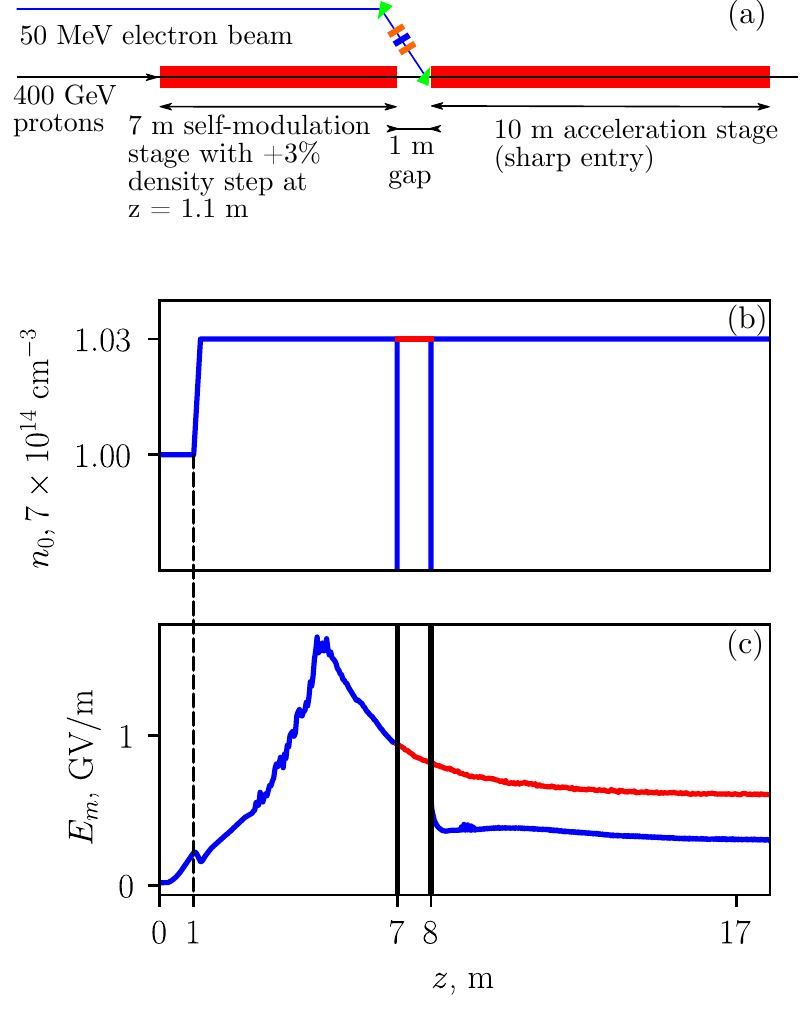}
\caption{A discussed scheme of AWAKE Run II: a general view (a), longitudinal dependencies of the plasma density $n_0$ (b) and wakefield amplitude $E_m$ (c) with (blue) and without (red) the vacuum gap.}\label{fig1-scheme}
\end{figure}

The effect was discovered when analyzing possible upgrades of the AWAKE experiment at CERN.\cite{NIMA-829-3,NIMA-829-76,PPCF60-014046} In this experiment, a long proton bunch undergoes seeded self-modulation in the plasma,\cite{PRL104-255003,PoP22-103110} splitting into short micro-bunches, which resonantly drive the plasma wave. During the first experimental run, the driver self-modulation \cite{Rieger,Turner} and witness  electron acceleration\cite{Wing} were demonstrated.
The second run (Run II) aims for high-quality of the accelerated electron bunch. One of the discussed Run II scenarios involves two plasma sections with a vacuum gap between them for injecting electrons (Fig.\,\ref{fig1-scheme}).\cite{IPAC16-2557,PRAB21-011301} The first section has a stepped-up longitudinal density profile for controlling the self-modulation.\cite{PoP18-024501} The created bunch train then enters the second section, excites a phase-stable wakefield there, and accelerates the electrons. In the considered scenario, the beam fluctuations manifest themselves particularly strongly, because the beam does not reach a perfect equilibrium in the first section and additionally deviates from the equilibrium state when passing through the vacuum gap. Driver density fluctuations lead to fluctuations of the focusing force, since the multi-bunch wave drive always operates in a weakly nonlinear regime\cite{PoP20-083119} and not in the blowout regime.\cite{PRA44-6189}

In Sec.~\ref{s2}, we describe how the emittance growth manifests itself in two-dimensional (axisymmetric) simulations of the AWAKE experiment. We identify the growth mechanism and discuss consequences of the axial symmetry. Then in Sec.~\ref{s3} we turn to the Cartesian model, which makes it possible to study beam loading effects. The beam loading turns out to be an effective way of reducing the emittance growth. In Sec.~\ref{s4}, we discuss the implications of the new effect.

\begin{table}[tb]
 \begin{center}
 \caption{Parameters for the AWAKE simulations.}\label{t1}
 \begin{tabular}{ll}\hline
  Parameter and notation & Value \\
  \hline
  \textbf{Proton driver:} \\
  Population, $N_b$ & $3\times 10^{11}$ \\
  Length, $\sigma_z$ & 6\,cm \\
  Radius, $\sigma_r$ & 160\,$\mu$m  \\
  Energy, $W_b$ & 400\,GeV \\
  Energy spread, $\delta W_b$ & 135\,MeV \\
  Normalized emittance, $\varepsilon_b$ & 2\,mm\,mrad \\
  \textbf{Plasma sections:} \\
  Length of the 1st cell & 7\,m\\
  Length of the vacuum gap, $L_g$ & 1\,m\\
  Length of the 2nd cell & 10\,m\\
  Plasma radius & 1.4\,mm\\
  Location of the density step & 1.1\,m\\
  Density before the step & $7\times 10^{14}\text{cm}^{-3}$\\
  Density after the step & $7.21\times 10^{14}\text{cm}^{-3}$\\
  \textbf{Witness bunch:} \\
  Length, $\sigma_{zw}$ & 10\,$\mu$m \\
  Radius, $\sigma_{rw}$ & 20\,$\mu$m  \\
  Energy, $W_w$ & 50\,MeV \\
  Delay relative to the laser pulse, $|\xi_w|$ \quad & $\approx 7.57$\,cm \\
\hline
 \end{tabular}
 \end{center}
\end{table}

\section{AWAKE simulations}\label{s2}

Since the baseline parameter set for the AWAKE Run~II is not decided at the time of our study, we take one of several discussed scenarios and additionally optimize it for the strongest established wakefield in the second section (Table~\ref{t1}, Fig.\,\ref{fig1-scheme}). The optimization consists in adjusting the location and magnitude of the plasma density step. We choose the 1\,m gap between the sections as a compromise between convenient injection of the witness bunch and wakefield reduction because of driver divergence in the gap.\cite{IPAC16-2557} As usual for AWAKE studies,\cite{NIMA-829-3} we assume that the plasma is instantly created by a short laser pulse co-propagating with the proton beam centroid. We also assume immobile ions and sharp plasma boundaries to exclude competing sources of emittance growth. Taking into account the transition regions\cite{JPD51-025203} would intermix the studied effect with a possible witness degradation during the injection process.\cite{PoP21-123116}

\begin{figure}[htb]\centering
 \includegraphics[width=0.95\columnwidth]{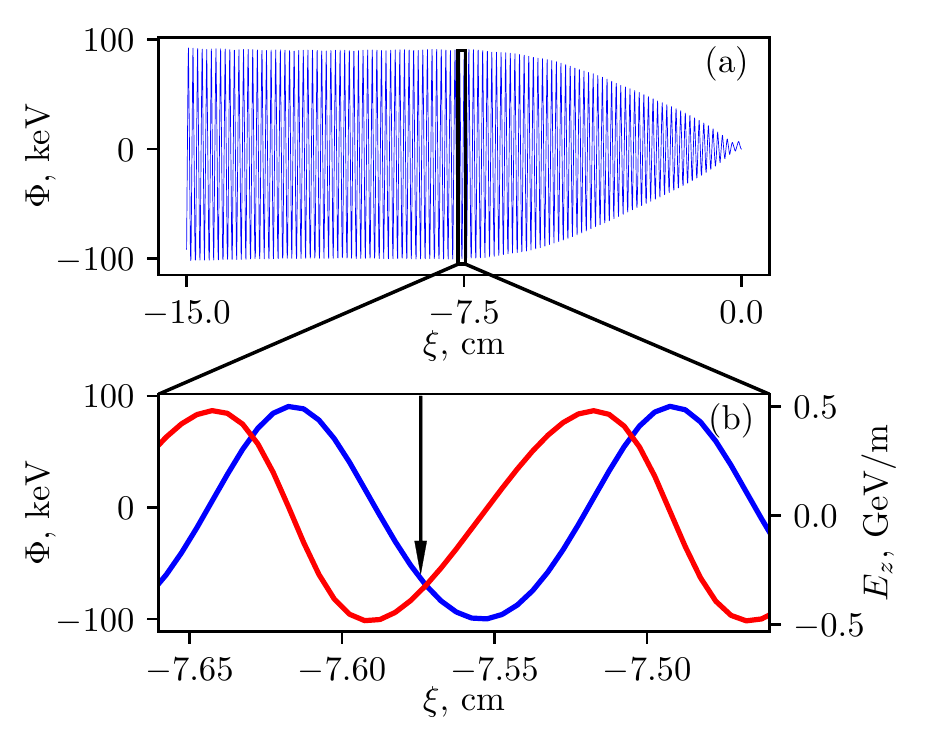}
\caption{The on-axis electric field $E_z$ (red line) and the wakefield potential energy $\Phi$ (blue line) at the beginning of the second section. Arrow shows the location of the test electron bunch.}\label{fig2-place}
\end{figure}

To study the quality of accelerating buckets, we inject small bunches of test electrons with zero energy spread and zero emittance to the places best suited for witness acceleration. The initial electron energy of 50\,MeV is sufficiently high to avoid longitudinal electron oscillations in the bucket,\cite{PoP21-123116} so we locally probe the wake in the cross-sections of injection. For the selected parameter set, these places are located about 60 wave periods behind the laser pulse [Fig.\,\ref{fig2-place}(a)], where the wave amplitude approaches its maximum, and at some fraction of the peak longitudinal field [Fig.\,\ref{fig2-place}(b)], where the witness is securely focused and the field profile can be flattened by loading a substantial witness charge.\cite{PRAB21-011301} All presented figures are for the cross-section at which the electron energy gain is 80\% of the maximum gain possible in this bucket.

We simulate the beam-plasma interaction with quasi-static particle-in-cell code LCODE.\cite{PRST-AB6-061301,NIMA-829-350} To avoid the emittance growth due to numerical effects, we use a fine simulation grid with radial and longitudinal steps $\Delta r = \Delta \xi = 0.005\, c/\omega_p = 1\,\mu$m, where $\xi = z-ct$ is the co-moving coordinate, $c$ is the speed of light, and $\omega_p$ is the plasma frequency. The longitudinal coordinate $z$ is measured from the entrance to the second plasma cell, and $t=0$ is the moment of laser pulse entry into the cell. The time step for the proton beam is $10\, \omega_p^{-1}$; it also determines the step $\Delta z = 10 \, c/\omega_p = 2$\,mm for calculating plasma fields in the quasi-static approach. The time step for low energy electrons is additionally reduced down to $7.8 \times 10^{-2} \omega_p^{-1}$. There are $1.2 \times 10^7$ equal macro-particles in the proton beam, about $3 \times 10^4$ weighted macro-particles in the electron witness, and 10 weighted plasma macro-particles per radial interval $\Delta r$.

\begin{figure}[htb]\centering
 \includegraphics[width=0.95\columnwidth]{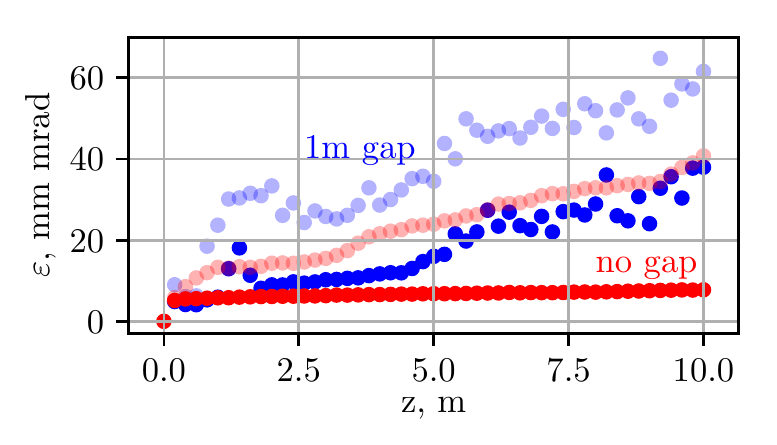}
\caption{Dependence of the normalized witness emittance $\varepsilon$ on the propagation distance in the second plasma cell $z$ in the presence of 1\,m vacuum gap between the cells (blue) and with no gap (red). Pale colors show results of lower-resolution simulations with $\Delta r = \Delta \xi = 0.01\, c/\omega_p = 2\,\mu$m, $\Delta z = 200\, c/\omega_p = 4$\,cm.}\label{fig3-effect}
\end{figure}
\begin{figure}[htb]\centering
 \includegraphics[width=0.95\columnwidth]{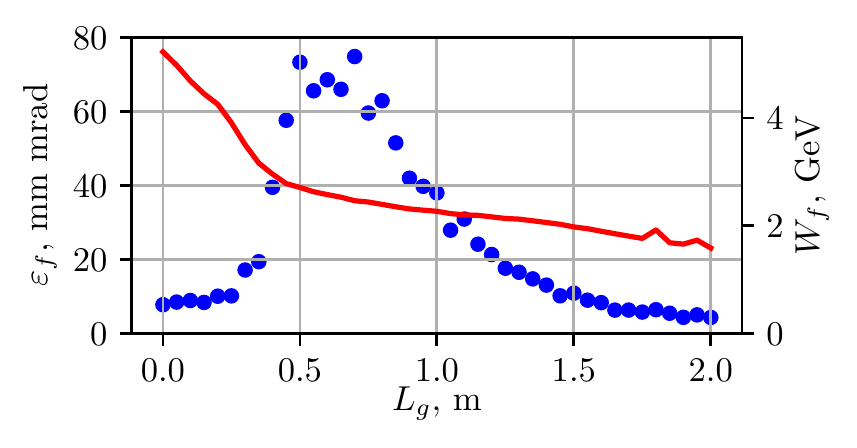}
\caption{Dependence of the final witness emittance $\varepsilon_f$ (blue points) and energy $W_f$ (red line) after propagating 10 meters in the plasma on the length $L_g$ of the vacuum gap between the plasma sections.}\label{fig4-gaps}
\end{figure}

Figure~\ref{fig3-effect} illustrates the discovered effect. At the very beginning of the second plasma cell, the normalized root-mean-square witness emittance $\varepsilon$ quickly reaches some equilibrium value of about 6\,mm\,mrad and then slowly grows if there is no gap between the cells. With the vacuum gap, however, the emittance grows much faster. This is a physical effect, as suggested by comparison with lower resolution runs. The emittance growth rate with no gap reduces as we increase the resolution, while with the gap it does not. Therefore we conclude that the growth rate with no gap gives us the upper limit on the contribution of numerical effects. These contributions are negligible in simulations of 1\,m gap case with the baseline resolution. The emittance $\varepsilon_f$ gained in the 10\,m long plasma section depends on the vacuum gap width (Fig.\,\ref{fig4-gaps}). As we see, 0.5\,m wide gap disturbs the drive beam in the most dangerous way, whereas the effect of wider gaps reduces disproportionately to the wave amplitude.

\begin{figure}[htb]\centering
 \includegraphics[width=0.95\columnwidth]{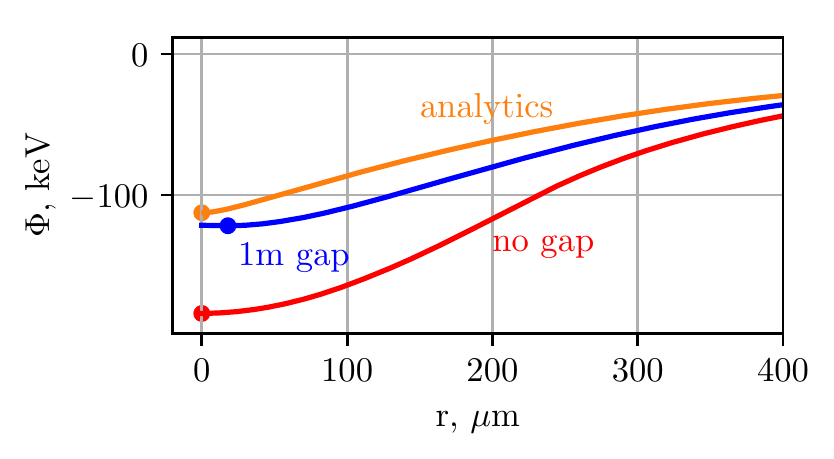}
\caption{Radial profiles of the wakefield potential energy $\Phi (r)$ experienced by the witness at $z=4$\,cm with and without the vacuum gap, and also calculated analytically according to the linear theory of plasma response for the case of 1\,m gap. The circles show the energy minima.}\label{fig5-potential}
\end{figure}

\begin{figure}[htb]\centering
 \includegraphics[width=0.95\columnwidth]{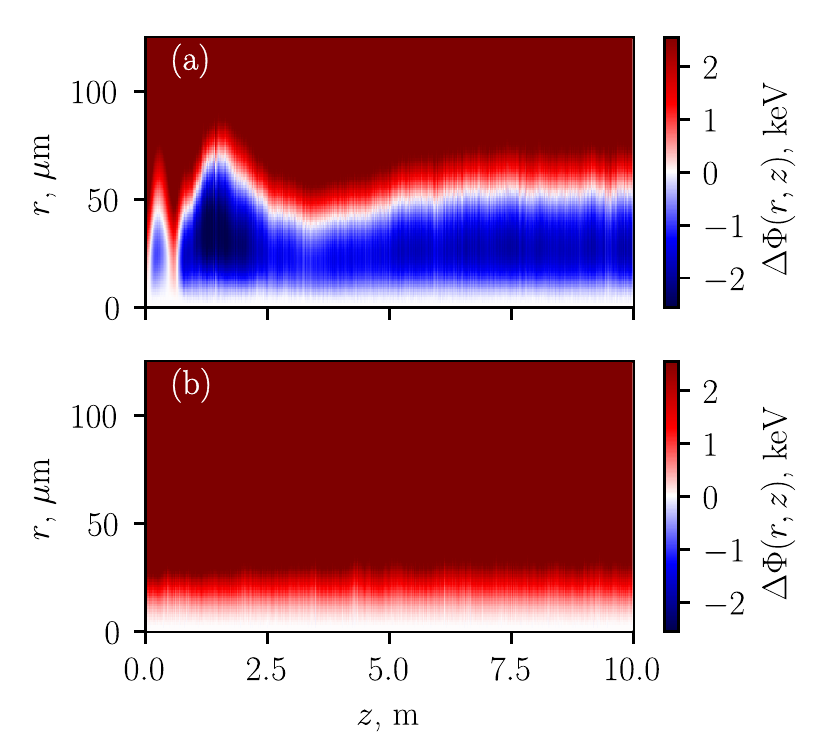}
\caption{Temporal variation of the potential energy in the near-axis region with (a) and without (b) the vacuum gap. For better visibility of the potential wells, the difference $\Delta \Phi (r,z) = \Phi (r, \xi_w, z) -\Phi (0, \xi_w, z)$ is shown.}\label{fig6-wells}
\end{figure}

The electron bunch gains emittance because of time-varying local maxima of the potential energy that appear on the axis [Figs.\,\ref{fig5-potential}, \ref{fig6-wells}(a)]. The wakefield potential energy $\Phi$ defines the force $\vec{F}$ acting on witness electrons, the components of which are
\begin{equation}\label{e1}
    F_\parallel = -e E_z = -\frac{\partial \Phi}{\partial z}, \quad
    F_\perp = -e (E_r - B_\phi) = -\frac{\partial \Phi}{\partial r},
\end{equation}
where $e>0$ is the elementary charge, and $\vec{E}$ and $\vec{B}$ are the electric and magnetic fields. A local potential hump defocuses witness electrons thus degrading the emittance. With no vacuum gap, there is always a potential well on the axis [Fig.\,\ref{fig6-wells}(b)], and the witness emittance is preserved.

\begin{figure}[htb]\centering
 \includegraphics[width=0.95\columnwidth]{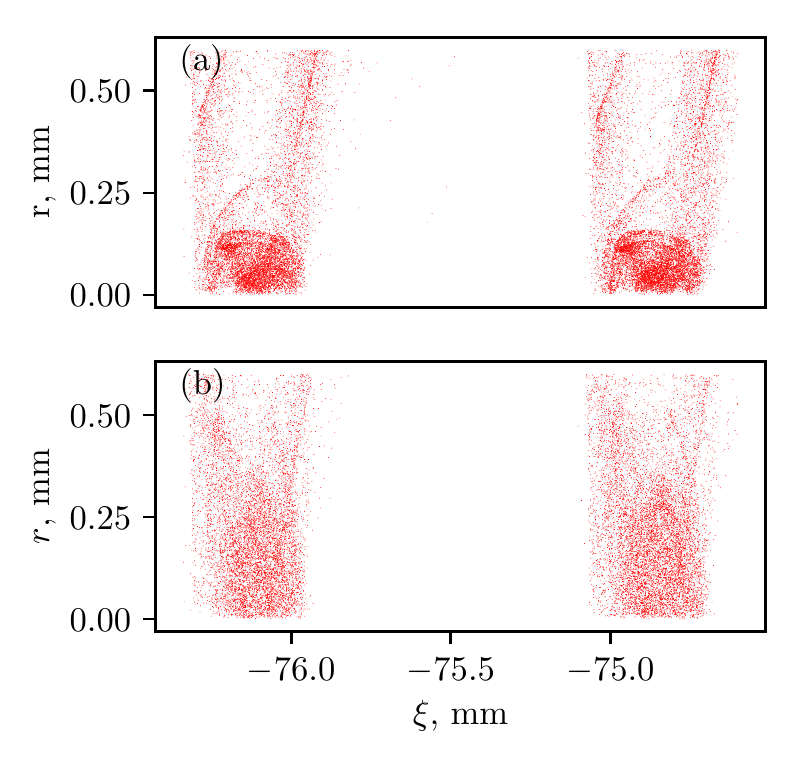}
\caption{A typical shape of the proton bunches before (a) and immediately after (b) the vacuum gap.}\label{fig7-portrait}
\end{figure}

The reason for the appearance of an unfavorable potential structure lies in nonlinear effects. With a strictly linear plasma response to the driver,\cite{PAcc20-171} an off-axis potential well is possible only with doughnut-shaped bunches that have a density dip on the axis at some cross-sections. No bunches of this kind were observed in simulations. The proton bunches radially expand in the gap (Fig.\,\ref{fig7-portrait}), and the change in their shape leads to the creation of a potential well with an almost flat bottom. This is a nonlinear effect, as the linear theory predicts a sharp potential minimum on the axis (Fig.\,\ref{fig5-potential}). Local fluctuations of the bunch density produce small additions to the wakefield potential, which form local maxima and minima against a background of approximately constant "bottom" level. These density fluctuations inevitably occur during self-modulation, are clearly visible in Fig.\,\ref{fig7-portrait}(a) and do not disappear after the vacuum gap, although they become less noticeable [Fig.\,\ref{fig7-portrait}(b)].

\begin{figure}[htb]\centering
 \includegraphics[width=0.95\columnwidth]{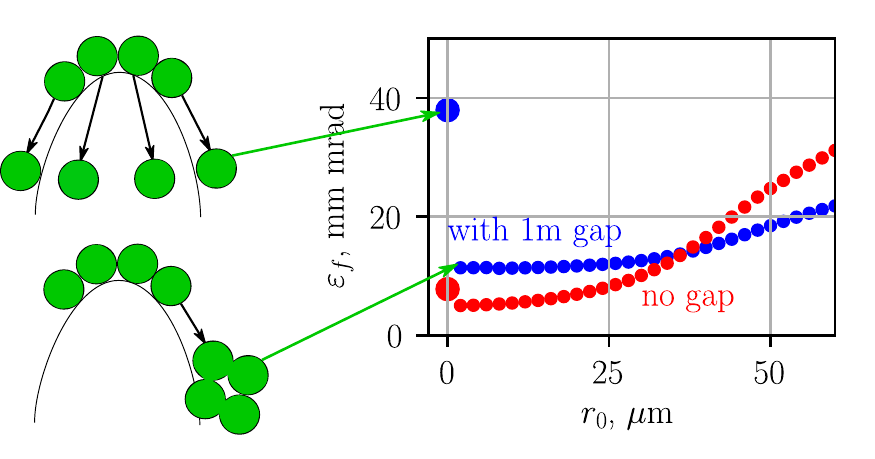}
\caption{Dependence of the final emittance $\varepsilon_f$ on the witness injection offset $r_0$ with and without the vacuum gap. The inset on the left illustrates two possible interpretations of the results of axisymmetric simulations.}\label{fig8-offax}
\end{figure}
\begin{figure}[htb]\centering
 \includegraphics[width=0.95\columnwidth]{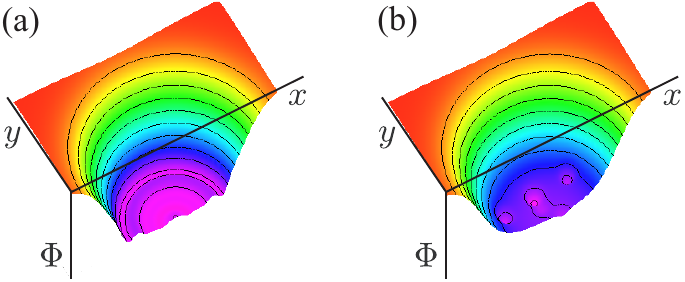}
\caption{Schematic representation of the wakefield potential energy at the witness position in the axisymmetric case (a) and expected in the 3D geometry (b).}\label{fig9-drawing}
\end{figure}

If the witness is injected collinearly but off-axis, the larger the offset $r_0$ is, the stronger the emittance increases regardless of the presence of a vacuum gap (Fig.\,\ref{fig8-offax}). The graph, however, contains a discontinuity at $r_0=0$, which comes from two different assumptions made about the azimuthal particle distribution of the witness. If we assume the electrons fall off the potential hump in an axisymmetric way, then the bunch size and emittance are larger. If all electrons fall to one side, then the emittance is lower. These two ways of interpreting the simulation results determine the limits within which the emittance varies in the real three-dimensional geometry. A three-dimensional analog of a non-stationary axisymmetric off-axis potential well [Fig.\,\ref{fig9-drawing}(a)] is a set of chaotically located potential wells and humps, size and position of which vary with time [Fig.\,\ref{fig9-drawing}(b)]. When the witness moves across this potential structure, both situations shown in the inset in Fig.\,\ref{fig8-offax} are possible, so the gained emittance will have an intermediate value.

\section{Plane case}\label{s3}

From the above discussion it follows that the effect of beam loading must be important for the emittance growth. If the witness is dense enough to create its own potential well, then the driver density fluctuations will have a weak effect on the motion of witness electrons, and the emittance will not rapidly grow. However the axisymmetric simulations do not account for the transverse displacements of the witness as a  whole (together with its local well). Three-dimensional simulations of this problem with the necessary temporal and spatial resolution are still beyond the capabilities of modern computing. Therefore, we move on to plane two-dimensional geometry.

\begin{figure}[htb]\centering
 \includegraphics[width=0.95\columnwidth]{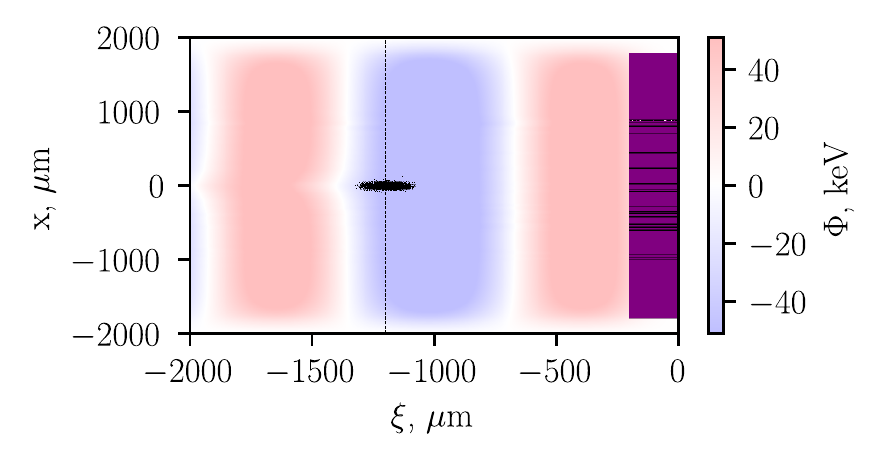}
\caption{The geometry of the plane problem: faint colors show the wakefield potential energy, the purple rectangle is the proton driver that has a uniform density distribution, dark bars on it are density fluctuations, black points are the witness electrons, and the vertical dashed line is the cross-section characterized in Fig.\,\ref{fig11-flat}.}\label{fig10-2d}
\end{figure}

Self-modulation of the proton beam in the plane geometry occurs quantitatively differently than in the axisymmetric case. Therefore, we do not simulate self-modulation, but reproduce the required potential behavior using a short driver with manually controlled density fluctuations (Fig.\,\ref{fig10-2d}). This approach allows us to formulate general conclusions about the dynamics of an electron bunch in a fluctuating potential, not limited solely to the AWAKE experiment.

\begin{figure}[htb]\centering
 \includegraphics[width=0.95\columnwidth]{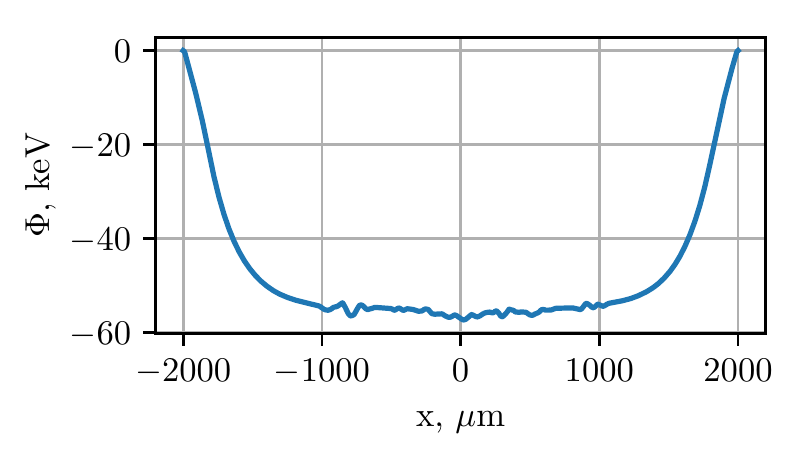}
\caption{The wakefield potential energy at the cross-section marked in Fig.\,\ref{fig10-2d} by the vertical dashed line in the absence of electron witness.}\label{fig11-flat}
\end{figure}
\begin{table}[htb]
 \begin{center}
 \caption{Beam and plasma parameters in the plane case.}\label{t2}
 \begin{tabular}{ll}\hline
  Parameter and notation & Value \\
  \hline
  \textbf{Driver:} \\
  Length, $l_{zb}$ & 200\,$\mu m$ \\
  Width, $l_{rb}$ & 3.6\,mm  \\
  Density, $n_b$ & $1.73\times 10^{13}\text{cm}^{-3}$ \\
  \textbf{Simulation area:} \\
  Window length & 2\,mm\\
  Window width & 4\,mm\\
  Beam propagation distance \quad & 10\,m\\
  Plasma density & $7\times 10^{14}\text{cm}^{-3}$\\
  \textbf{Witness:} \\
  Length, $\sigma_{zw}$ & 60\,$\mu$m \\
  Radius, $\sigma_{rw}$ & 34\,$\mu$m  \\
  Peak density, $n_w$ & $2\times 10^{14}\text{cm}^{-3}$ \\
  Energy, $W_w$ & 50\,MeV \\
  Energy spread, $\delta W_e$ & 0 \\
  Normalized emittance, $\varepsilon$ & 2\,mm\,mrad \\
\hline
 \end{tabular}
 \end{center}
\end{table}

We compose the wakefield potential energy of two parts: a stationary flat-bottom well and time-dependent small perturbations against its background (Fig.\,\ref{fig11-flat}). The first part is created by a wide proton bunch with a uniform density distribution. The bunch density is such as to provide the same acceleration rate (250 MeV/m) as in the axisymmetric case (Table~\ref{t2}). The influence of plasma fields on this bunch is turned off. The second part is created by small localized time-dependent perturbations to the drive beam. Each elementary perturbation has the form
\begin{equation}\label{e2}
    \delta n_b (x) = \begin{cases}
    n_f, & |x-x_c|<x_f/3, \\
    -n_f/2, & x_f/3<|x-x_c|<x_f, \\
    0, & \text{otherwise}.
    \end{cases}
\end{equation}
The location $x_c$ of the perturbation is random and uniformly distributed along the transverse coordinate $x$ in the the interval $|x_c| < 1\,\text{mm}$. Density perturbations of this shape do not change the average energy level and produce localized potential energy perturbations of the same transverse size $x_f$. Therefore, we can directly control the size of small potential wells and choose it in accordance with the results of axisymmetric simulations [Fig.\,\ref{fig6-wells}(a)], i.e., about $10\,\mu$m. At any moment, there are 20 elementary perturbations \eqref{e2}, 10 short-living and 10 long-living ones. Short-living perturbations appear at random places for the time period of $200\omega_p^{-1}$ that corresponds to 4\,cm of beam propagation. Long-living perturbations do not change locations, but their amplitudes $n_f$ vary proportionally to $|\sin [2 \pi (t-t_0)/\tau_f]|$ with $\tau_f = 4000\omega_p^{-1} = 80\,\text{cm}/c$ and random $t_0$. The exact values of amplitudes $n_f$ are chosen to match the emittance growth rate observed in axisymmetric simulations (Fig.\,\ref{fig3-effect}). The witness parameters (Table~\ref{t2}) are are chosen to simulate the mode, where the witness creates a deep potential well, but does not reach the complete blowout.

\begin{figure}[htb]\centering
 \includegraphics[width=0.95\columnwidth]{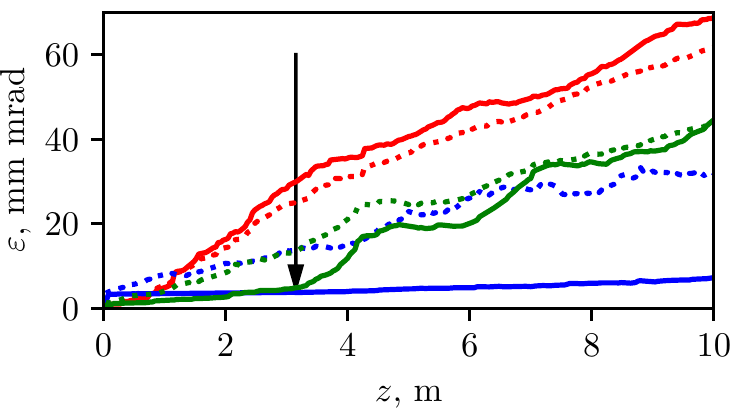}
\caption{Dependence of the witness emittance $\varepsilon$ on the propagation length $z$ for different witness charges: the full charge as in Table~\ref{t2} (blue lines), 10\% of the full charge (green lines), and test electrons with a negligible charge (red lines). The dotted lines denote the emittance of the entire bunch, and the solid lines represent the emittance of the central $40\mu$m-long slice marked in Fig.\,\ref{fig13}. The arrow marks arrival of the erosion front to the central slice of the partially charged beam.}\label{fig12}
\end{figure}

\begin{figure}[htb]\centering
 \includegraphics[width=0.95\columnwidth]{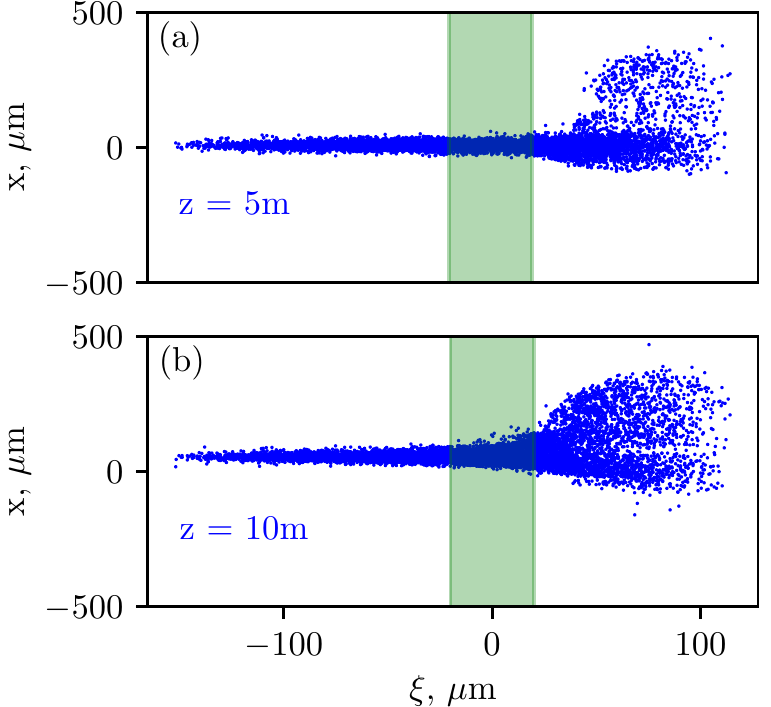}
\caption{Portraits of the  full-charge witness after propagating 5\,m (a) and 10\,m (b) in the plasma. The colored rectangles mark the witness slice, the emittance of which is shown in Fig.\,\ref{fig12} by a solid line.}\label{fig13}
\end{figure}
The plane case gives us an insight into the influence of the witness charge on the emittance growth (Fig.\,\ref{fig12}). An increase in the beam charge reduces the emittance growth rate. For bunches with nonzero charge, the emittance of the central fragment grows much more slowly than the emittance of the beam as a whole. This can be explained by rapid destruction of witness head (Fig.\,\ref{fig13}), which is not confined by the self fields and fully experiences the potential fluctuations. They gradually ruin the beam emittance from head to tail.

\section{Discussion}\label{s4}

While analyzing one of the upgrade options for the AWAKE experiment, we discovered the problem of witness emittance growth due to fluctuations of the focusing force. The problem is applicable to all plasma wakefield acceleration schemes with linear or moderately nonlinear plasma waves. Such schemes have recently become popular again, as they offer the advantage of symmetric acceleration of electrons and positrons\cite{AIP1299-3,PRST-AB13-101301,PRST-AB14-091301,PRST-AB15-051301} and allow the use of positively charged drivers.\cite{Nat.524-442,RAST9-63,RAST9-85} The importance of the discovered effect will increase as plasma acceleration techniques will approach collider applications and the requirements to the witness quality will become more stringent.

The blowout,\cite{PRA44-6189} or bubble\cite{APB74-355} regime is immune to this mechanism of emittance growth, as the focusing force in the bubble is fully determined by the ion background and does not fluctuate with time. The quasi-nonlinear regime,\cite{PRL88-014802,PRST-AB7-061302,AIP1299-500} in which the witness is only partially residing in the bubble, however, may be subject to the emittance growth, as the witness head will experience the focusing force fluctuations.

The main cause of the time-dependent transverse force is a driver-plasma mismatch. The equilibrium state of a particle bunch in its own wakefield is rather exotic and strongly differs from the usual Gaussian distributions in coordinates and momenta.\cite{PoP24-023119} Therefore, any driver will change its shape after entering the plasma, thus creating a time-varying wakefield. Even with the exactly matched beam radius, some equilibration of the beam shape will still occur. Laser pulses may also produce a non-stationary wave, if mismatched to the focusing channel.\cite{RMP81-1229} Therefore, the beginning of the plasma section, where the wakefield fluctuations are strongest, is the most dangerous for the witness quality. Perhaps a witness injection from the side\cite{JPP78-455} after the driver reach the radial equilibrium, is free of the above effect, but this has yet to be investigated.

The emittance growth rate decreases with increasing the witness charge due to additional focusing of the witness by its own wakefield. Nevertheless, the witness head always degrades. There is a clearly visible boundary between eroded and intact parts of the witness, which slowly propagates backward along the bunch. Quantitative characteristics of witness erosion, however, depend on the particular setup, so we describe the erosion process only qualitatively.

\acknowledgments

This work is supported by The Russian Science Foundation, grant No.~14-50-00080. The computer simulations are made at Siberian Supercomputer Center SB RAS.

\end{document}